# Evaluating Granularity in Markov Chain-Based Trust Models for Vehicular Ad Hoc Networks (VANETs)

Rezvi Shahariar

1.  Institute of Information Technology, University of Dhaka, Dhaka, Bangladesh

**Abstract**: Trust management is a critical research pillar in Vehicular Ad Hoc Networks (VANETs), where the reliability of shared data depends entirely on driver integrity. In these networks, a driver's reputation is dynamically constructed based on the veracity of their recent message history: consistent reliability builds trust, while frequent misinformation leads to exclusion. This study analyses driver announcement characteristics by modelling behavioural transitions—specifically the frequency and nature of shifts between "good" and "bad" states. To facilitate this analysis, three distinct Markov chain-based behavioural models are evaluated with varying degrees of granularity: a 4-state model, a 7-state model, and a high-resolution 11-state model. By simulating announcement and reporting patterns, each model's ability to reflect nuanced behavioural shifts is assessed. Our results confirm that increasing the number of trust states significantly enhances the system's ability to capture complex, dynamic driver behaviours, providing a more robust framework for security in VANETs.

## INTRODUCTION

Vehicular Ad Hoc Networks (VANETs) help drivers and other road users obtain real-time traffic information. Drivers may plan their journeys by considering several factors, for example, distance, road smoothness, and low traffic. These considerations help drivers reach their destinations with comfort. However, while driving, incidents may happen and cause trouble for others. If an event is not announced promptly, it causes more traffic chaos. In VANETs, Roadside Units (RSUs) and the Trust Authority (TA) are the two authoritative units, while private and public vehicles are the users of VANET. Figure 1 demonstrates the components of a VANET. In a VANET, one vehicle announces a message (traffic event) which others relay in their vicinity. Messages are disseminated by the participating vehicles in the VANET. Most of the time, RSUs announce information about the current traffic situation of a particular road. So, when a user vehicle gets a message from other vehicles or RSUs, it decides whether to use a particular part of a road. To make this decision, the vehicle evaluates the trust of the received message. In [1], the authors perform a comparative analysis of using both false and true message announcements in two alternate route scenarios. They measure the average travel time of all vehicles from one point to another during the simulation. Their results suggest that both false messages and unannounced true messages impact a driver's journey time. From these results, trustworthy announcements are always needed in VANETs. Otherwise, vehicles form queues around an incident on a road.

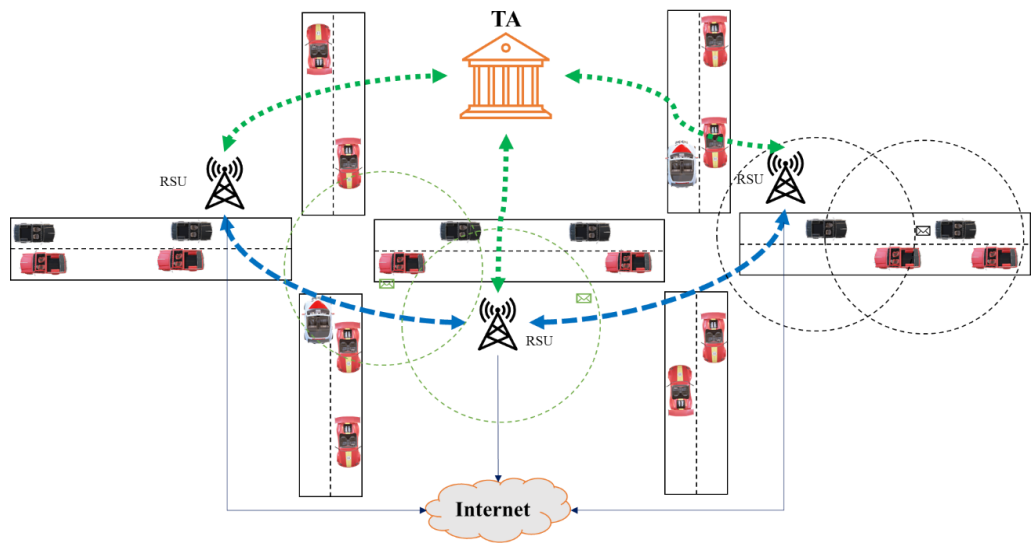

**Figure 1**: A usual VANET scenario

The trust of a message depends on several factors, such as message freshness, temporal proximity, role, experience, honesty, forwarding capability, the driver's previous history of message announcements, and the driver's current trust. There are three kinds of trust evaluation procedures: data-oriented, entity-oriented, and hybrid. Data-oriented trust evaluation mainly considers several metrics related to the incident, for example, message freshness, time closeness, and distance. Entity-oriented trust evaluation evaluates the trust of the entity, for example, a driver or a vehicle. This kind of evaluation considers the driver's trust, role, experience and so on. Finally, the hybrid approach evaluates both data and entity trust. For example, this evaluation may consider message freshness and driver current trust to find the trustworthiness of an incident.

In VANETs, many types of traffic events are so common, for example, accidents, traffic jams, stranded vehicles, diversions, road closures and so on. [2] presents a traffic event management framework considering some traffic events, i.e., accidents, traffic jams, diversion. This framework illustrates the whole sequence of operations to sort out some events from the road. Usually, traffic patterns change when an event happens in an area. Chaos is also seen when drivers announce false messages, and other drivers believe the offender's message to meet his/her goal. When many drivers believe the wrongdoer mal-intent message, then traffic congestion may happen due to the diverted traffic on a specific route.

The driver's message announcement behaviour needs to be examined so that I can identify when a driver announces a trustworthy or untrustworthy message. To investigate this problem, I have developed three distinct Markov chain process-based driver-behaviour models. This model is used to capture the interactions between different drivers in terms of the likelihood of trustworthy announcements from different trust states. Each model has different trust states, and drivers switch trust states based on their announcements. If the vehicle announces trustworthy messages repeatedly, then it moves to higher trust states, from where the probability of getting trustworthy messages is increased. Alternatively, vehicles, when announcing more untrustworthy messages, move into untrusted states from where they tend to send more untrustworthy messages. Consequently, they move to lower trust states. From a lower trust state, they announce more untrustworthy messages unless they try to be honest. If they remain untrusted, they may be blacklisted as soon as their trust reaches an unacceptable trust level set by the trust model. Nonetheless, some vehicles which may improve their trust score even from lower trust states by announcing only trustworthy messages. To simulate this behaviour, the drivers from lower-trust states can announce true events with lower probability. An opposite characteristic may be noticed in some drivers when they belong to higher trust states. Some drivers may lose trust by announcing untrue messages several times from higher trust states. For each state, different probabilities for both trustworthy and untrue message announcements are set. Three different types of activities for simulating these behaviours are considered, which are announcements, reporting an announcement, and clarifying an event to RSU queries. Though the trust model considered does not add a reward for the clarifier vehicle, which sends a response when an RSU collects data from the neighbourhood. Using these driver behaviour models, one can simulate different behaviour from drivers by setting different probabilities.

Section II reviews recent work on behaviour analysis based on different parameters. In Section III, two new driver behaviour models are presented. After that, Section IV depicts the driver behaviour analysis of drivers using three different Markov chain-based driver behaviour models. Finally, in Section V, this work is concluded by highlighting the contributions of this work.

## RELATED WORK

Many trust models have been proposed in the literature, and they used many different kinds of algorithms, for example, blockchain, machine learning, fuzzy, and so on, for evaluating the trust of drivers. Additionally, many used a Markov chain or Hidden Markov model to analyse driver behaviour using steering and brake data. However, only a few in the literature perform Markov chain-based driver behaviour analysis based on trustworthy announcements. Additionally, many surveys are conducted to present the current achievements, identifying future research directions for trust management in VANETs. [3] presents a survey which classifies trust model follow either the sender side or the receiver side evaluation. This also suggests that the trust model must follow sender-side evaluation to achieve fast decision capability to avoid a probable traffic jam.

In [4], the authors propose a trust model which verifies trust at the sending side using a Tamper Proof Device to secure trust. This model uses response time and distance as the key metrics to assign reward or punishment by the trust manager of drivers. However, when a dispute happens, only feedback is collected to verify the event by the RSU. Additionally, a fuzzy reward or punishment evaluation technique is also proposed in [5]. This evaluation considers the driver's history, the severity of the event, and the RSU confidence and appropriately assigns a reward or punishment to the driver.

[6] proposes a Hidden Markov Model-based driver behaviour recognition mechanism using braking and steering data. This avoids collision on the road using a Hidden Markov Model-based actuation behaviour model. Additionally, this employs Dempster Shafer Theorem to address the uncertainty issue in VANETs. In [7], researchers consider smartphones as a data collection method from drivers to observe driver action, distraction, and response time. The Multiple Markov Switching Variable Auto-Regression (MSVAR) model is used to fit the gathered behavioural data. This model can be used in safety applications.

[8] collects driver behaviour data using the smartphones of the driver and the client. This analyses this data using Long Short-Term Memory (LSTM), Convolutional LSTM, and Convolutional Neural Networks (CNN-LSTM). This model considers slow, normal, and aggressive behaviour of the driver. This system is implemented in a ride-sharing application to inform customers about driver behaviour in advance. In [9], a behavioural analysis is conducted to show whether a driver is aggressive or not using supervised machine learning algorithms (Support Vector Machine, Logistic Regression, and Random Forest) and deep learning algorithms (Multiple Layer Perceptron and Recurrent Neural Network). This research collects speed, acceleration, fuel consumption, and direction data from the built-in sensors of vehicles.

In [10], a survey on different driver behaviour models is presented. Since VANET is a heterogeneous environment, this considers both human driven vehicle and autonomous vehicles as elements of VANET. This shows the importance of capturing human driver behaviour in different contexts. Thus, an autonomous vehicle should predict the actions of a human driver in VANETs. This survey gives importance on car following model and the lane changing characteristic. [11] presents a sequential deep neural network to estimate reward based on the communication of a driver. A similar model is used to classify drivers as fraudulent versus non-fraudulent. This model computes trust from the earned reward of drivers. In [12], researchers investigate the trust and experienced driver behaviour. This also isolates the trusted driver using six different driving scenarios. The scenarios include cruising, curve, intersection, lane-switch, cut-in, and lane addition/reduction. This research suggests that trust should account for the functionality of autonomous vehicle driving, driving experiences, and driving contexts. In [13], driver behaviour analysis is performed using neural networks. This model identifies different driving styles using multiple factors, for example, ecological traits, road and vehicle situations, event class and recognition, as well as natural and running conditions.

In the literature, the behavioural analysis of trustworthy announcements based on a Markov chain model is not used widely. In my previously published work in [5], I consider only one Markov state transition diagram for announcement behavioural analysis of drivers. Then I think I should consider multiple driver behaviour models having varying trust states. I would like to record the impact of varying trust states on the announcement characteristics using different Markovian state transition diagrams. From this analysis, I would like to find out the advantages of additional trust states in the driver behaviour model. In the next section, two new driver behaviour models and one existing driver behaviour model from [5] are discussed.

## MARKOV STATE TRANSITION-BASED DRIVER BEHAVIOUR MODEL

In a vehicular environment, vehicles announce a traffic incident when they encounter one on the road. There are many different types of traffic incidents. Some of them occur frequently, while others occur occasionally. Vehicular announcement helping others to detour to a specific road to avoid the road containing a traffic incident. When a driver announces a true event, he/she builds up trust from it. As other drivers find this is a helpful activity, they treat the event announcement positively. So, they consider the sender more trustworthy. Alternatively, when the same driver announces a false event, drivers get annoyed if they detour and later find the event was fraudulent. In some situations, a driver may become trustworthy for a while, and at other times he/she may turn into untrustworthy. Thus, this type of behaviour needs to be examined to see when a driver sends more untrustworthy messages. If this can be modelled in a Markov chain-based driver behaviour model, having different trust states with a defined probability of sending trustworthy and untrusted announcements from all trust states. It is necessary to consider many different driver behaviour models having a varying number of trust states. Hence, three different types of Markov chain-based driver behaviour models are considered.

**Driver Behaviour Model with Four Trust States**

Figure 2 shows a Markov chain-based driver behavioural model which consists of four different trust states. The trust states are "Blacklisted", "Bad", "Normal" and "Good". Different probabilities are assigned for both sender and reporter drivers. Here, sender drivers announce an event. Alternatively, reporter drivers report an event from a sender driver as they believe the event is false. They can do this both truly and maliciously. The reporting of an untrue attack is the activity of complaining to an RSU that the announced event is deemed to be false to a reporter driver. Clarifier drivers send feedback about an event when an RSU asks for clarification for an event. Their replying pattern can also be set using different probabilities. In the real world, they would as they notice on the road. After that, RSU determines the trustworthiness of the event by applying a weighted voting-based approach. In this process, RSU collects both feedback and drivers' trust to find the sum of the product of trust and feedback. If this sum is greater than zero, then the RSU detects that the originator sends a trustworthy message, and the reporter announces an untrue attack. If the sum is less than zero, then reward and punishment are given differently. RSU allocates reward and punishment to corresponding drivers based on this decision. It is expected that drivers will announce more trustworthy announcements when they announce from "Good" and higher-trusted states. Hence, in Figure 2, the probability of sending trustworthy messages from S3 is set to higher values than from S2 and S1. Table 1 defines the probability of sending trustworthy and untrusted messages from each trust state.

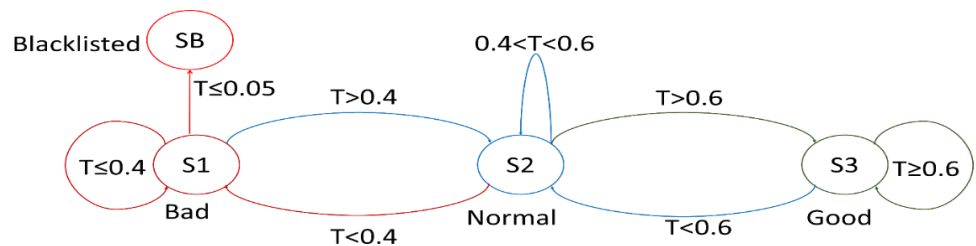

**Figure 2**: Markov chain-based driver behaviour model with only four trust states

Table 1: Announcement Probabilities

| Trust State | Probability of announcing trustworthy messages from the sender driver | Probability of announcing untrue messages from the sender driver |
|---|---|---|
| SB | 0 | 0 |
| S1 | 0.40 | 0.60 |
| S2 | 0.50 | 0.50 |
| S3 | 0.80 | 0.20 |

**Driver Behaviour Model with Six Trust States**

Figure 3 shows another driver behaviour model that is used in the driver behaviour analysis in [5]. This paper also works with the trust model presented in [4]. As stated in Figure 3, the driver behaviour model consists of six different trust states, which are "Blacklisted", "Very Bad", "Bad", "Normal", "Good" and "Very Good". Vehicles cannot make any announcements from a "Blacklisted" state as set by the trust framework in [4]. Only a few types of incidents can be announced from the "Very Bad" and "Bad" states. In "Normal" trust state, vehicles can announce most traffic events. From "Good" and "Very Good" trust state, all traffic events can be announced. At startup, vehicles start their journey from "Normal" state and based on their behaviour of announcements and reporting of attacks, they either build or lose trust and change trust states accordingly. From the "Normal" trust state, when trust T is greater than 0.6, drivers enter the "Good" state; otherwise, they move into the "Bad" trust state when trust T is less than 0.49. This is how drivers move from one trust state into another trust state. The probability distribution of true and untrue announcements from each state is defined in Table 2 for the sender vehicle. Note that different behaviour can be simulated using the different probability distributions for both sender and reporter vehicles from each trust state.

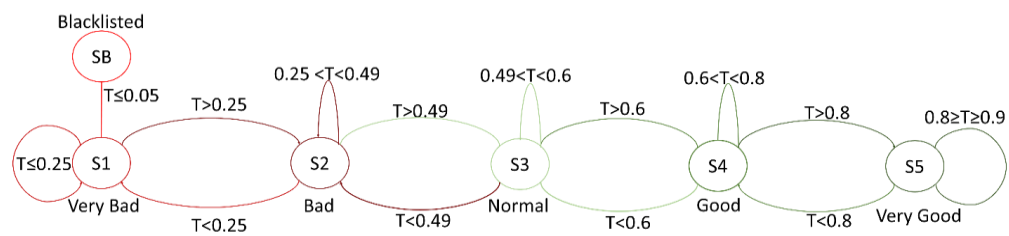

**Figure 3**: Markov chain-based driver behaviour model with six states [5]

**Table 2**: Probability distribution of true and untrue announcements [5]

| Trust States | Probability of Announcing a Trustworthy Message | Probability of Announcing a Malicious Message |
|---|---|---|
| "Very Good" | 0.8 | 0.2 |
| "Good" | 0.6 | 0.4 |
| "Normal" | 0.4 | 0.6 |
| "Bad" | 0.2 | 0.8 |
| "Very Bad" | 0.1 | 0.9 |
| "Blacklisted" | 0 | 0 |

**Driver Behaviour Model with Eleven Trust States**

In this driver behaviour model, a few additional distinct trust states beyond those specified in Figures 2 and 3 are added. The states are "Blacklisted", "Very Bad", "Bad", "Fairly Bad", "Below Normal", "Normal", "Above Normal", "Fairly Good", "Good", "Very Good", and "Outstanding". In this model, the trust range is divided into many smaller trust ranges, and for each range of values, one trust state is added to the model. The trust range defines the boundary of trust values to remain at each state. Many trust states can be used to differentiate announcements of events based on their severity. For example, some events like a bush on the road, a narrow road alert, which can be announced by drivers from low-trust states. On the other hand, severe kind of events should be announced by highly trusted drivers associated with high trust states. The probabilities of trustworthy announcements and untrue announcements are set differently for all states, as shown in Table 3. With a higher trust value, drivers tend to announce or report more accurately, whereas from a lower trust state, they tend to report more fraudulently. Probability is set for all states using this consideration. Different probabilistic distributions for all states can be used when different characteristics of drivers need to be tested.

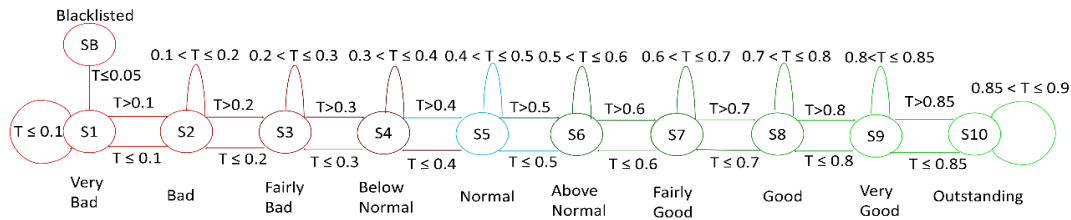

**Figure 4**: Driver behaviour-based trust model with eleven trust states.

**Table 3**: Probability of true and untrue announcements

| Trust states | Probability of announcing trustworthy messages | Probability of Announcing Untrue Messages |
|---|---|---|
| "Blacklisted" | 0 | 0 |
| "Very Bad" | 0.1 | 0.9 |
| "Bad" | 0.2 | 0.8 |
| "Fairly Bad" | 0.3 | 0.7 |
| "Below Normal" | 0.4 | 0.6 |
| "Normal" | 0.5 | 0.5 |
| "Above Normal" | 0.6 | 0.4 |
| "Fairly Good" | 0.7 | 0.3 |
| "Good" | 0.8 | 0.2 |
| "Very Good" | 0.85 | 0.15 |
| "Outstanding" | 0.95 | 0.05 |

## PERFORMANCE EVALUATION

Driver behaviour models shown in Figures 2, 3, and 4 are implemented in Veins [14], which consists of OMNeT++ and SUMO. Inside Veins, SUMO is a traffic simulator, and OMNeT++ works as a network simulator. Both simulators communicate through TRaCI (Traffic Control Interface). Veins is used for simulating the aforementioned driver behaviour models, as it is freely available. Also, the Erlangen road network scenario from Veins is used. A long circular route is created where all vehicles travel until the simulation ends. Vehicles can move on roads in both directions using the car following model. Vehicles are kept constant in all experiments. A traffic flow containing 100 vehicles is used in all experiments. In this series of experiments, initial trust of vehicles differs widely, which supports the trust framework in [4]. Vehicle changes trust based on the driver's behaviour, which is decided by the RSU whether the driver states truly or falsely. An RSU allocate reward or punishes drivers based on the dispute resolution process illustrated in [4]. In this series of experiments, only an accident on a road is announced, which carries the accident information on a particular road section. There is only one vehicle which announces event periodically. Though in real life, any vehicle can face an accident on a road, and then it announces that to others. The simulation is conducted this way to track the trust changes for a particular sender and many deterministic reporters. The reporters are deterministic in the sense that they would either report or not report an event based on their probability distribution of the trust states they belong to. A reporter vehicle from lower-trust states would report more announcements falsely, whereas from higher-trust states, only a few announcements would be reported fraudulently. The experiments are carried out in the presence of twelve RSUs and one TA (Trust Authority). RSU continually monitors the network based on the traffic incidents collected from the neighbouring vehicles. It would finish the rebroadcasting only when the announced event is sorted.

This research investigates whether a trust model with many states is a better model than a trust model with fewer states. With driver behaviour models, announcement credibility varies based on the probability distribution. Thus, two additional variants of the trust-based driver behaviour model with eleven and four trust states are considered. In [5], only one driver behaviour model with six trust states is specified, and the analysis is performed using only this one. After conducting a series of experiments, results suggest that a driver behaviour model with higher trust states can capture more dynamics in driver announcement behaviour. Table 4 lists the simulation parameters for experimenting with driver announcement behaviour.

**Table 4: Experimental Parameters**

| Parameters | Values |
|---|---|
| Trust update | RSU reward and punishment |
| Simulation period | 5000 seconds |
| Warm-up period | 400 seconds |
| Announcement intervals | 500 seconds |
| Initial trust | Variable (0.5 to 0.9) |
| Vehicle numbers | Constant (100) |
| Number of RSUs | 12 |
| Transmission range | 300 meters |
| Presence of a TA | Yes |
| Attacker model | Untrue and Inconsistent attack |
| Clarifying reward | Fixed (0.08) but not added |
| Collaboration Timer | 120 seconds |
| Area | Urban |
| Number of lanes on the road | Two lanes |
| Use of Car Following Model | Yes |

**Scenario 1: Using the Markov Driver Model with Four States**

In the next few experiments, I have used the following probabilistic distributions for reporter drivers. Table 5 shows that from the S1 state, a reporter driver truly announces 60% of the time, whilst falsely announces 40% of the time.

Table 5: Probability distribution for reporter drivers

| Trust state | Reporter's Probability of Announcing Untrue Messages | Reporter's probability of not announcing untrue messages |
|---|---|---|
| SB | 0 | 0 |
| S1 | 0.60 | 0.40 |
| S2 | 0.5 | 0.50 |
| S3 | 0.2 | 0.8 |

*With a 0.7 Trust Score for the Sender Driver and 0.6 for Reporter Drivers*

In this experiment, all vehicles start their journey with a 0.6 trust score except the sender vehicle, which starts with a 0.7 trust score. They start their journey from a "Good" trust state, from where 80% announcements should be trustworthy. Here, V1, V2, V3, V4, and V5 are reporting the announcements from V0 based on their probability. In Figure 5, x-axis shows the simulation period, whereas the y-axis shows the trust changes of the driver.

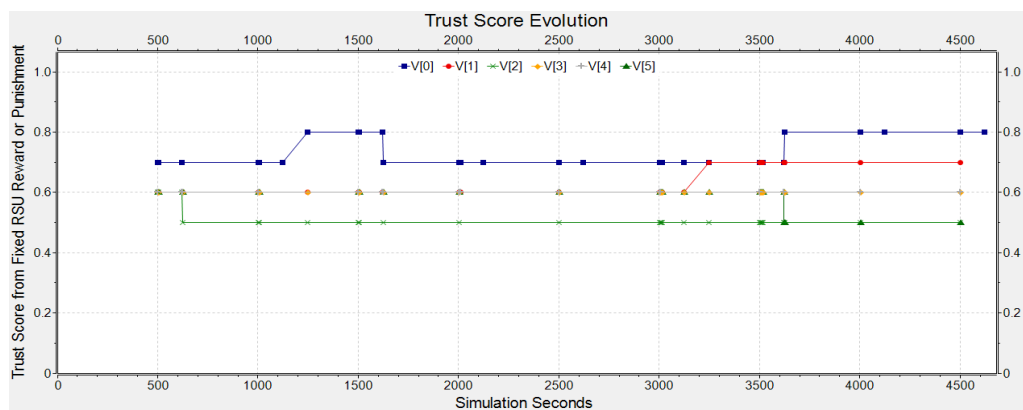

Figure 5: Trust score changes of sender and reporter drivers

A warmup period (400 seconds) is noticeable as no event is recorded during this period. After elapsing this period, trust scores are recorded during the entire simulation. Although V2 reports an incident fraudulently with a trust score of 0.6 and receives RSU punishment using an untrue detection procedure, which reduces its trust score. After this, V2 enters "Normal" trust state. As V0 wins this dispute, it builds more trust from this reporting. The event at 1000s is not reported, so no trust increment or decrement happens. For the event announcement at 1500s, V3 reports the event from V0, and V3 builds its trust. V0 loses trust at 1700s, but it remains in "Good" trust state. After this event, no other events are reported until the 3500s. At this point, V5 announces an untrue report, which results in punishment from the RSU, and V0 earns an RSU reward and sets its trust to 0.8 and finishes with this trust score. V5 receives RSU punishment at the same time and enters into "Normal" trust state from "Good" trust state.

### With a 0.6 Trust Score for the Sender Driver and 0.7 for Reporter Drivers

In this experiment, V0 starts with a 0.6 trust score, and all other vehicles start their journey with 0.7. At 500 seconds, V1 reports an announcement from V0, but later receives the punishment from the resolver RSU. In this case, V0 improves trust from this dispute as it wins where most clarifiers also believe that V0 announces a true accident event. Then, an event at 1000s is also reported by V2, and hence V0 builds trust at 2100s, and V2 loses trust at around 1600s. The event at 1500s is also reported by V2, for which V0 improves trust again and reaches the peak and V2's trust is reduced further. Furthermore, the announcements at 2000s and 2500s are reported by V5 and at 3000s are reported by V2. These disputes cause trust loss for both V2 and V5, but V0's trust remains the same, as it already hit the peak. In this experiment, all vehicles remain in "Good" trust state except the V2 and V5, which enter "Normal" trust state at 2120s and 3120s respectively. Figure 6 shows this scenario, where simulation seconds are shown on the x-axis and trust changes are shown on the y-axis.

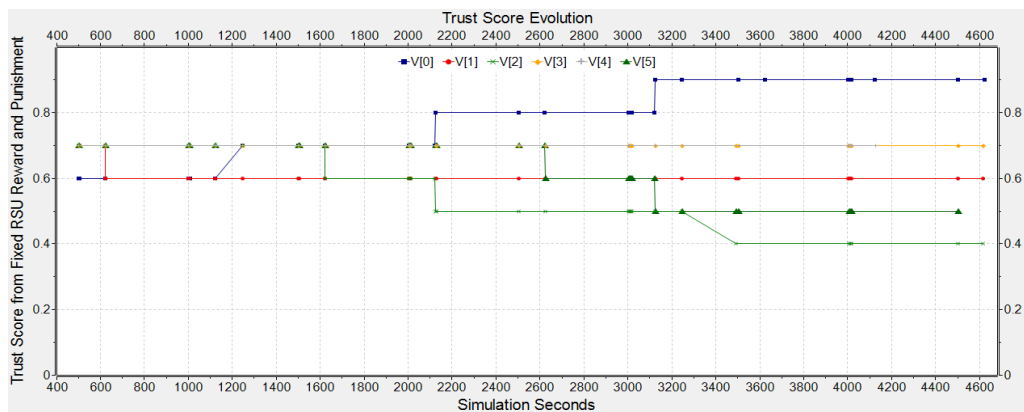

**Figure 6**: Trust score changes of sender and reporter drivers

### With a 0.9 Trust Score for the Sender Driver and 0.8 for the Reporter Drivers

In this experiment, I set the initial trust to the highest value for V0, and other vehicles start their journey with 0.8. For this experiment, most vehicles are trusted in the network, and their trust states are "Good".

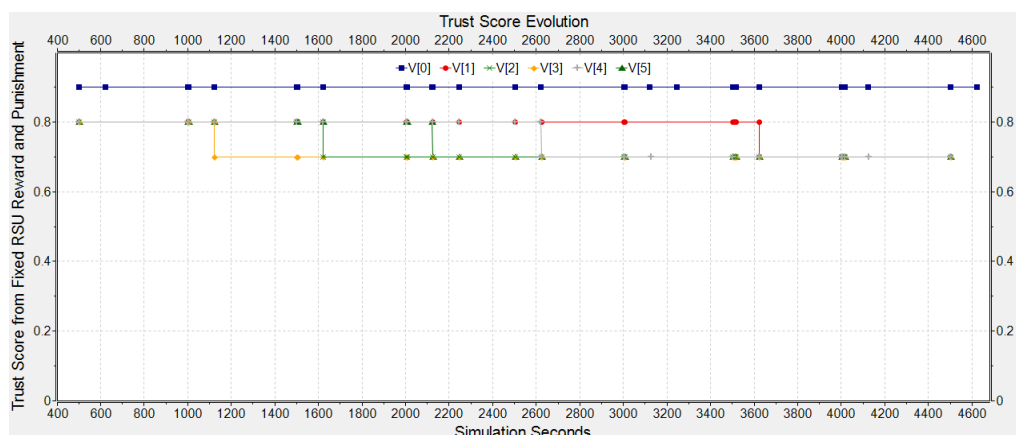

**Figure 7**: Trust score changes of sender and reporter drivers

I did not start from other trust states, as I announced only a severe traffic incident, which does not allow for a "Bad" trust state. As V0 is trusted, there are more chances that V0's announcements will be trustworthy, which is also proved when disputes arise. This is why V0's trust remains constant through this experiment. On the other hand, all reporters were proven to be mischievous, though they are highly trusted. This simulates the behaviour that a highly trusted vehicle can also announce untrue events, as the probability of sending untrue reporting from "Good" trust state is 0.2. Thus, these untrue reports are proved at the resolver RSU. Their trust reduces after the dispute resolution process resolves all disputes. Figure 7 shows this scenario, where simulation seconds are on the x-axis and trust changes are on the y-axis.

*With a 0.8 Trust Score for the Sender Driver and 0.9 for Reporter Drivers*

Figure 8 depicts vehicles starting their journey from a "Good" trust state, but results in a different characteristic from the above. In this experiment, though a trusted vehicle reports at 500s, it is proven untrue at an RSU. Thus, V0 receives a reward, whereas the reporter V1 receives a punishment. This has also happened on other occasions, for example, 1000s, 1500s, from V1. V2 and V4 also report trustworthy announcements fraudulently at 2000s and 2500s, and in the 3500s. This results in their trust reduction. This behaviour proves the fact that a trusted driver can report untrue reports or announcements. Similarly, a low-trusted driver can build trust by turning its behaviour trustworthy.

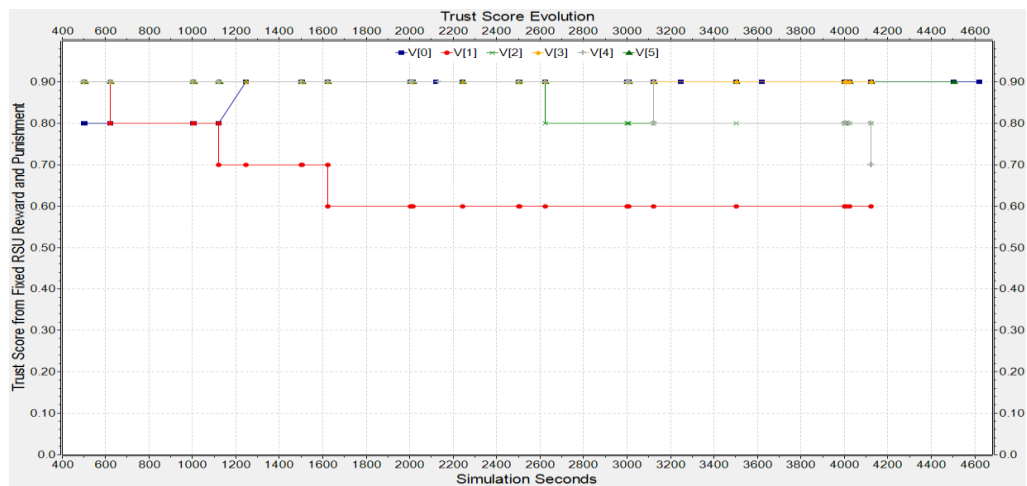

**Figure 8**: Trust score changes of sender and reporter drivers

## DISCUSSION ON RESULTS

While these four trust curves provide valuable insights, they fail to capture a critical nuance: significant fluctuations in trustworthiness can occur within a single "Good" trust state. Currently, some drivers maintain a "Good" rating despite a drop in trust value, while others reach the upper limits of the same category.

This lack of granularity stems from the overly broad range assigned to this specific trust state. To address this and achieve a more precise driver behavior model, a higher resolution of trust states is required. Consequently, the following experiments evaluate the model presented in [5] alongside a newly proposed model, illustrated in Figure 4.

**Scenario 2: Using The Markov Driver Model With Six Trust States**

In the next few experiments, I have used the following probabilistic distributions for reporter drivers. Table 6 says that from the "Very Good" state, a reporter driver truly announces 90% of the time, whilst falsely announces 10% of the time.

**Table 6**: Probability distribution for true and untrue reporting [5]

| Trust States | Probability of Reporting Fraudulent | Probability of trustworthy Reporting |
|---|---|---|
| "Very Good" | 0.1 | 0.9 |
| "Good" | 0.3 | 0.7 |
| "Normal" | 0.5 | 0.5 |
| "Bad" | 0.7 | 0.3 |
| "Very Bad" | 0.9 | 0.1 |
| "Blacklisted" | 0 | 0 |

*With a Trust Value of 0.5 for all Drivers*

I have used the same Markov driver behaviour model presented in [5]. In this experiment, all vehicles start their journey with a trust score of 0.5 and from a "Normal" trust state. The announcement and reporting probability are also taken from [5]. V5 first reports the announcement of V0. It is proven false to the resolver RSU, so it assigns punishment to V0, and V5 gets the RSU reward.

After this, no announcement was possible for V0 as its trust reduces to 0.4 and it moves to "Bad" trust state. The trust framework in [4] does not support any kind of announcement from this state. This happens as I only announce one severe type of incident, so the access control mechanism limits the scheduled announcements. As a result, all vehicles finish their journey with these trust scores. Figure 9 shows this scenario, where simulation seconds are shown on the x-axis and trust changes are shown on the y-axis.

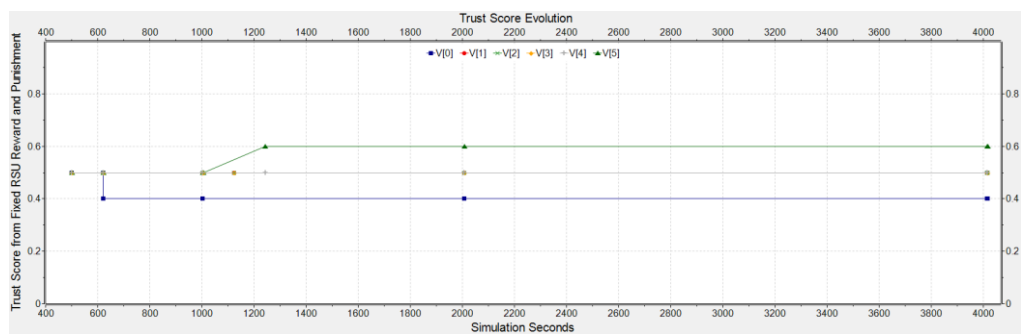

**Figure 9**: Trust score changes of sender and reporter drivers

*With a Trust Value of 0.6 for all Drivers*

In this experiment, all vehicles start with a trust of 0.6 with a "Good" trust state. At 500s, a vehicle V5 reports an untrue announcement from V0. V5 wins the dispute, and it builds trust. V0 loses trust in this announcement and moves to "Normal" trust state. Vehicle V3 also reports an announcement from V0 at 1500s, for which it gets punishment and is moved into "Normal" trust state, as it reports an untrue attack falsely. V0 builds trust from this dispute at 2100s and again moved back to "Good" trust state. V0 also develops trust at the same time for the report generated from V2 at 2000s. It is noticed that these two rewards are added almost at the same time. Next, an announcement at 3000s is reported by V3 again, which is false and moved into "Bad" state, so it receives punishment at 3120s, and V0 receives RSU reward at the same time and moved to "Very Good" trust state. The next announcement at 3500s is reported, which is proven false to an RSU, and it punishes V0. But the reporter vehicle does not receive the reward as the environment is wireless, where any packet may collide with other packets or when the RSU broadcasts the reward message, the reporter was not present in the vicinity of the RSU. This is also true for the incident at 4000s, which causes V1 to receive punishment. But V0 misses the RSU reward due to a similar reason. However, this can be saved by an RSU when a vehicle enters its zone, which then broadcasts the message. Then reward or punishment from RSU will not be lost. Also, I switched off the beaconing, so the RSU cannot tell when a specific vehicle enters and leaves its region. But in real life, entering an RSU's area can be detected from the beacon message of a vehicle, and then the allocated reward or punishment can be relayed to the vehicle. Figure 10 shows this scenario, where simulation seconds are shown on the x-axis and trust changes are shown on the y-axis.

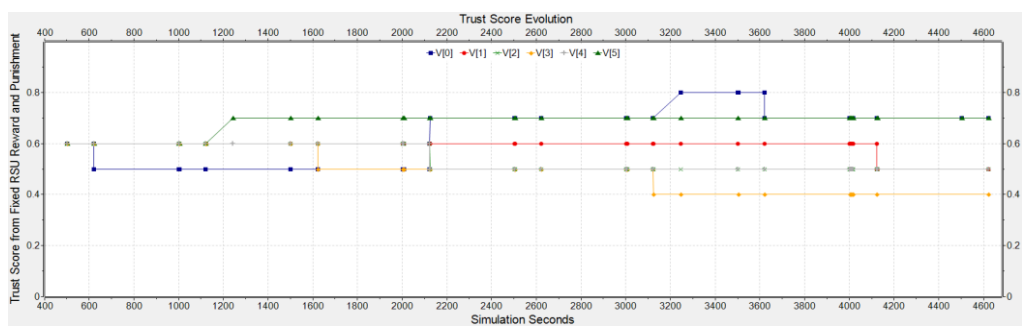

**Figure 10**: Trust score changes of sender and reporter drivers

*With a Trust Value of 0.6 for the Sender Driver and 0.5 for all Drivers*

V0 starts from a "Good" trust state with a 0.6 trust score, and all vehicles start with a 0.5 trust score from a "Normal" trust state. The announcement at 500s is reported by V5. V5 receives RSU punishment and moves into "Bad" trust state.

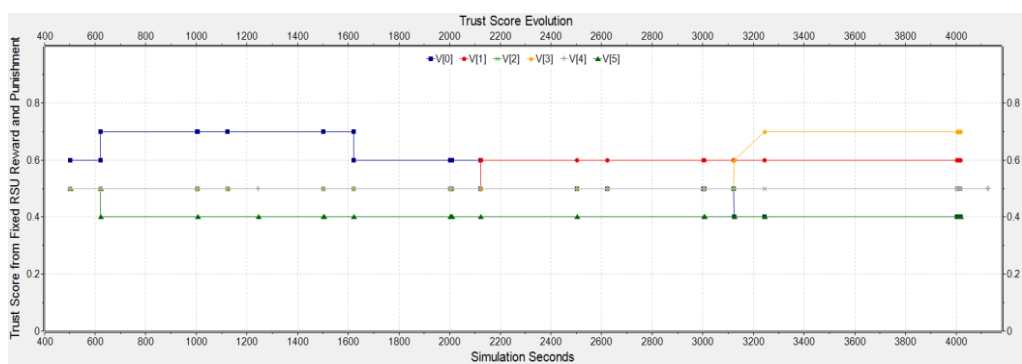

**Figure 11**: Trust score changes of sender and reporter drivers

On the other hand, V0 receives the RSU reward at 620s. The announcement at 1500s is reported by V1, which builds up trust, and V0 receives RSU punishment for announcing an untrue event. The event at 2000s is also reported by V3. V3 receives RSU reward at 3100s, and V0 receives RSU punishment at 2120s for the untrue event reporting, and V0 entered into "Normal" trust state. V3 again reports an untrue attack for the announcement at 3000s. Once again, this proves to be an untrue announcement from V0 for which it receives punishment at 3120s. V0 now enters into "Bad" state. V3 receives the RSU reward a bit later at 3320s. After that, no attack reporting takes place in this experiment as V0 sends nothing from the "Bad" trust state. Figure 11 shows this scenario, where simulation seconds are shown on the x-axis and trust changes are shown on the y-axis.

*With a Trust Value of 0.8 for the Sender Driver and 0.7 for all Drivers*

The chart in Figure 12 shows the trust changes among the vehicles, where V0 starts with 0.8 from "Very Good" trust state, and all other vehicles start with 0.7 trust score from "Good" trust state. V0's first announcement is reported by V5 falsely, and its trust declines. Then, the announcement of 1000s is reported by V2, for which it receives a reward, which is added at 2100s and entered into "Very Good" state. This untrue reporting by V2 causes V0 to receive punishment at 1620s. The announcement of V0 at 1500s is reported by V4, for which it receives the RSU reward at 2120s. V4 receives punishment at the same time and enters into "Good" trust state. The announcements reported at 2000s and 2500s are fraudulent, and V3 received RSU punishment and moved into "Normal" trust state. As V0 reaches the peak trust, its trust is not improved. V0's announcement of the 3000s is also reported by V4, for which it obtains a reward from RSU. The punishment for V0 is not received at the V0 due to a collision, or when the RSU was broadcasting, and V0 was far from the broadcasting location. I also limit the relaying to 3 hops. The latter announcements of V0 at 3500s and 4000s are also reported by V4 and V5 in turn, and their reporting was proved false to an RSU for which they received RSU punishments at 3620s and 4120s. As a result, V5 moved into "Normal" trust state. Figure 12 shows this scenario, where simulation seconds are shown on the x-axis and trust changes are shown on the y-axis.

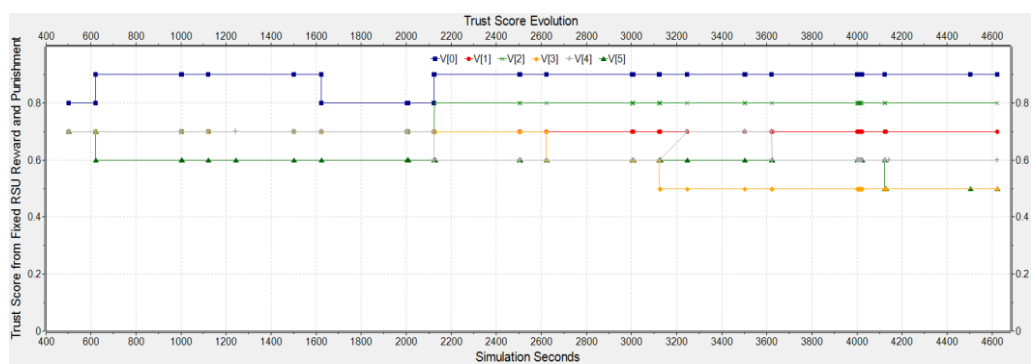

**Figure 12**: Trust score changes of sender and reporter drivers

*With a Trust Value of 0.85 for the Sender Driver and 0.8 for all Drivers*

This experiment shows a trust increment or decrement when V0 starts with 0.85, and all other vehicles start with 0.8. Here, all vehicles start from a "very Good" trust state. As V0 is highly trusted, it broadcasts trustworthy messages only. When any vehicle reports an untrue attack fraudulently, these are disproved at the respective RSUs, and hence their trust curves show a small decline by the amount of RSU punishments. This curve in Figure 13 also shows another pattern that trusted vehicles can also announce or report fraudulently. Because of this, their trust decreases, and they move into lower trust states. For example, V3 enters into "Bad" trust state and reaches 0.4 trust score from where it cannot send untrue reports or severe traffic incidents, though it starts from "Very Good" trust state with 0.8 trust score. Figure 13 shows this scenario, where simulation seconds are shown on the x-axis and trust changes are shown on the y-axis.

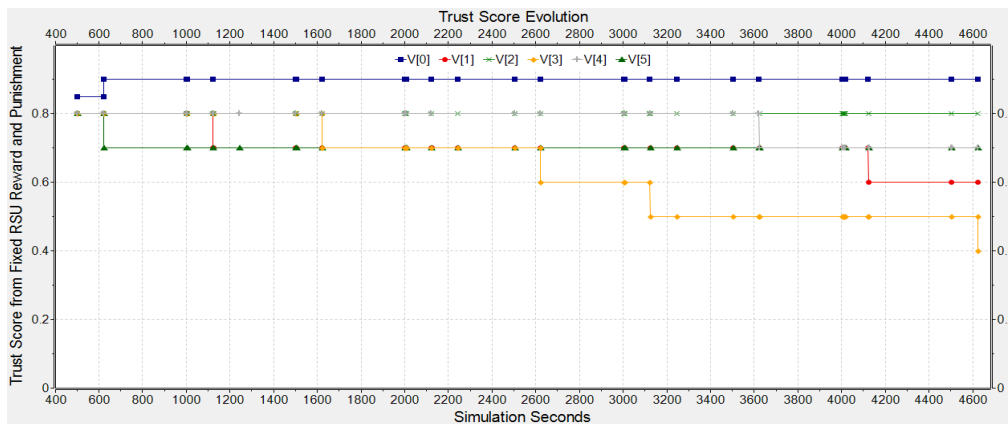

**Figure 13**: Trust score changes of sender and reporter *drivers*

*With a Trust Value of 0.9 for the Sender Driver and 0.85 for all Drivers*

Vehicle V0 starts with a 0.9 trust score from a "Very Good" trust state, and other vehicles start with a 0.85 trust score from the same trust state. 0.9 is the highest trust score permitted by the trust framework presented in [4]. Vehicles V4, V5, and V3 reports few announcements, and these are disproved at RSUs. Reporter vehicles announce fraudulently with a probability of 0.1 from this trust state. These vehicles receive RSU punishments, which reduce their trust score. As V0 only announce trustworthy messages, it only receives RSU rewards, which are received by V0, but it is not added by V0's trust manager. As the framework supports a maximum trust score of 0.9, V0's trust is not changed throughout its journey. Figure 14 shows this scenario, where simulation seconds are shown on the x-axis and trust changes are shown on the y-axis.

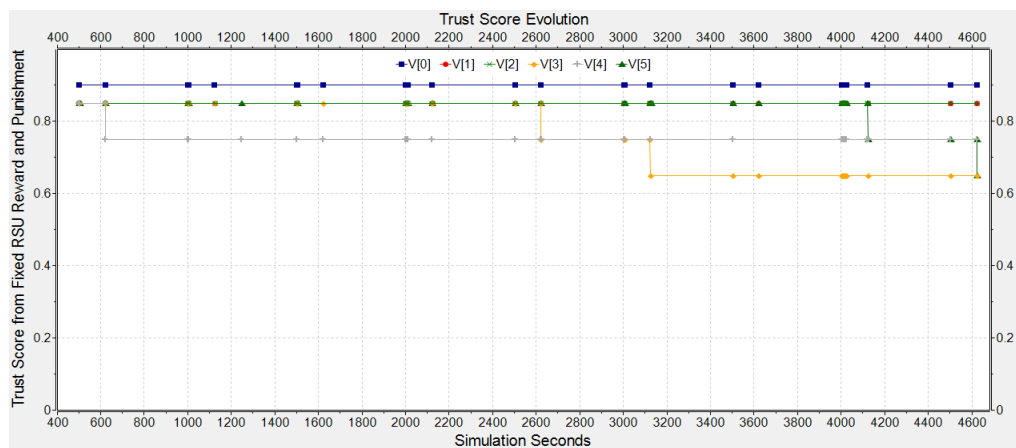

**Figure 14**: Trust score changes of sender and reporter drivers

*With a Trust Value of 0.9 for all Drivers*

This experiment also shows the trust characteristics of sender and reporter vehicles when all vehicles start their journey with 0.9. Two vehicles, V3 and V5 reports announcements of V0, which clarifiers clarify as true to RSUs. Hence, they receive RSU punishments, which reduce their trust value and change their trust states to lower trust states. From a low trust state like "Normal or Bad", I set a higher lying probability to tell a lie or to report a trustworthy announcement fraudulently. Hence, these two, especially V5 reduces trust in the reporting of trustworthy announcements of V0. Figure 15 shows this scenario, where simulation seconds are shown on the x-axis and trust changes are shown on the y-axis.

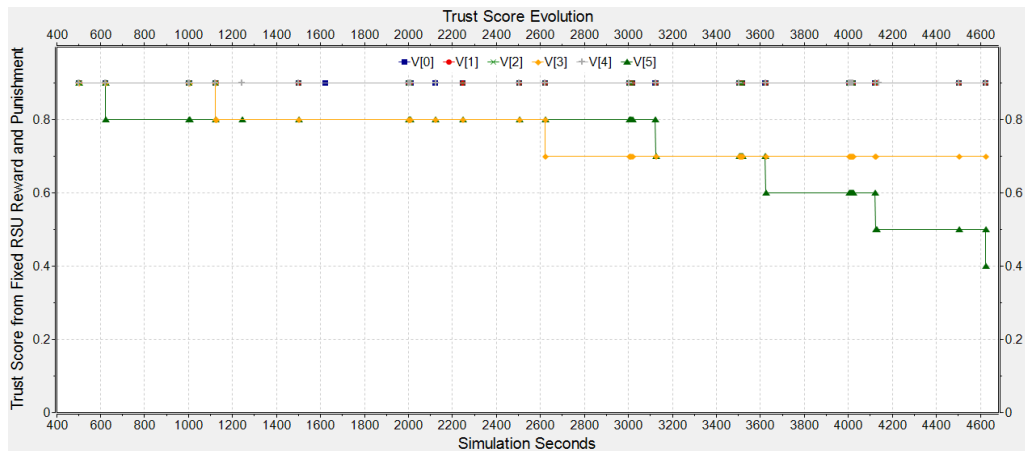

**Figure 15**: Trust score changes of sender and reporter drivers

## DISCUSSION ON RESULTS

In this set of experiments, driver behaviour is tested when their trust lies between 0.5 and 0.9. For the analysis, only three states are used, which are "Normal", "Good", and "Very Good". During the simulation, it is found that some drivers in the 1$^{st}$ and 3$^{rd}$ experiments announce untrue events more, and they receive RSU punishments. On the other hand, in experiments 4, 5, 6, and 7, sender drivers build trust throughout the simulation period. In experiments 5 and 7, some reporter drivers show very poor behaviour, whereas in all experiments, both good and bad reporters are present. Consequently, reporters exhibit varying trajectories, either gaining or losing trust throughout the simulation. The behaviour analysis performed using the state transition-based model in Figure 3 is good, but it is better to explore a bit more by introducing additional trust states in the next scenario.

**Scenario 3: Experimenting with Driver Behaviour with Eleven States Markov-Chain Based Driver Behaviour Model**

In the next few experiments, I have used the following probabilistic distributions for reporter drivers. I have used the Markov chain-based driver behaviour model in Figure 4. It has eleven trust states which are "Blacklisted", "Very Bad", "Bad", "Fairly Bad", "Below Normal", "Normal", "Above Normal", "Fairly Good", "Good", "Very Good", and "Outstanding".

Table 7 says that from "Normal" state, a reporter driver truly announces 45% of the time whilst falsely announce 55% of the time.

Table 7: **Probability of true and untrue announcements from all states**

| Trust states | Probability of false reporting | Probability of trustworthy reporting |
|---|---|---|
| "Blacklisted" | 0 | 0 |
| "Very Bad" | 0.9 | 0.1 |
| "Bad" | 0.8 | 0.2 |
| "Fairly Bad" | 0.7 | 0.3 |
| "Below Normal" | 0.6 | 0.4 |
| "Normal" | 0.55 | 0.45 |
| "Above Normal" | 0.5 | 0.5 |
| "Fairly Good" | 0.4 | 0.6 |
| "Good" | 0.3 | 0.7 |
| "Very Good" | 0.2 | 0.8 |
| "Outstanding" | 0.1 | 0.9 |

*With a Trust Value of 0.9 for all Drivers*

In this experiment, the trust score of all vehicles is set to 0.9, which belongs to the "Outstanding" trust state. V4 reports the event at 2000s, which is proved later as true reporting. So, V4 receives a reward, and V0 gets a punishment at 2120s and enters into "Very Good" trust state. As V4 has a peak trust value, its trust state is not changed. The announcement at 3000s is reported by V5, but the respective RSU disproves it. V5 gets RSU punishment, which takes it into "Very Good" state, and V0 again reaches peak trust, then enters "Outstanding" state. The announcement at 3500s is reported again by V4, and it gets RSU punishment, which reduces its trust by 0.1, and it enters the "Very Good" state. V4 reports another transmission from V0 at 4000s, which was a false transmission from V0. After that, V4 enters into "Outstanding" trust state from "Very Good" state. However, the curve does not show any trust change due to collision, or V0 was distant from the location where the punishment was broadcasted. This is common in the wireless environment. Figure 16 shows this scenario, where simulation seconds are shown on the x-axis and trust changes are shown on the y-axis.

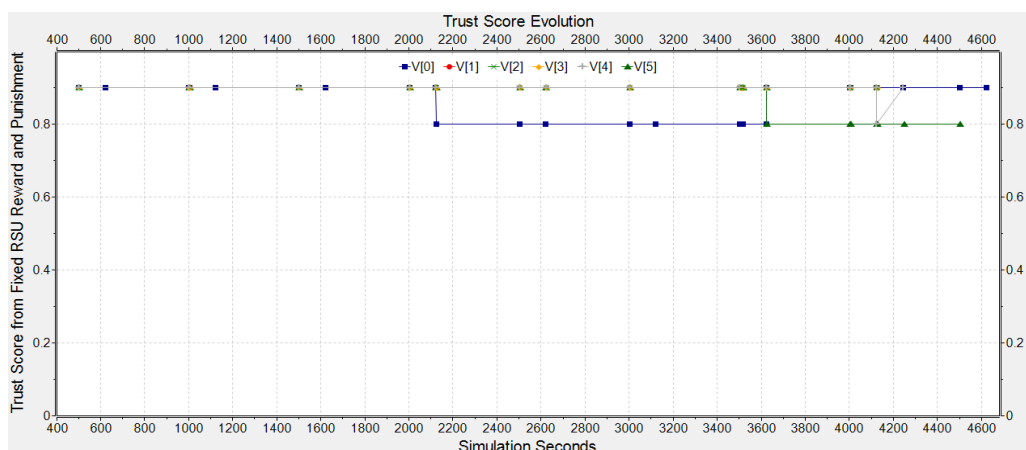

**Figure 16:** Trust score changes of sender and reporter drivers

*With a Trust Value of 0.8 for all Drivers*

In this experiment, all vehicles start their journey with a 0.8 trust score from a "Good" trust state. V1 reports an announcement of V0 at 1500s, which disproves at the RSU. Then, V4 loses trust, and it enters into "Fairly Good" state. V0 builds trust from this dispute and enters into the "Outstanding" state. Later, the announcement at 2000s is also reported by vehicle V2, which gets RSU punishment and enters into "Fairly Good" state. V0's trust does not change as its trust touches its peak. Then, V0's next announcement at 2500s is reported by V1. Like before, V1 face the same result as it reports V0's announcement fraudulently. Then V1 enters the "Above Normal" state. V2 reports the announcement at 3000s fraudulently, for which it receives the same punishment from the RSU at 3120s and enters into "Above Normal" state. There are no further reporting activities noticed after the announcement at 3000s. Figure 17 shows this scenario, where simulation seconds are shown on the x-axis and trust changes are shown on the y-axis.

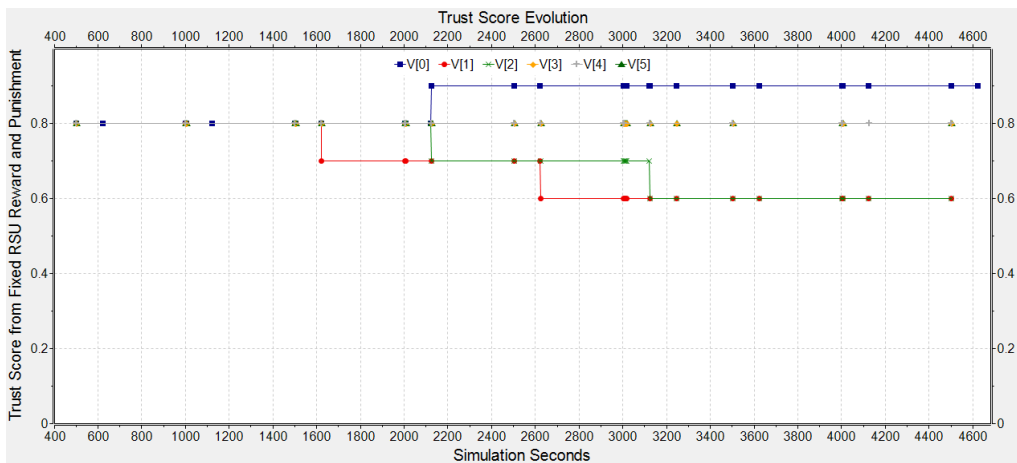

**Figure 17:** Trust score changes of sender and reporter drivers

*With a Trust Value of 0.8 for the Sender Driver and 0.9 for all Drivers*

The announcements from V0 are reported at 1500s, 2000s, 4000s, and at 4500s. As most vehicles start with a higher trust state, "Outstanding", so very few reports are generated fraudulently compared to the previous experiments. V0 is seen as trustworthy and enters into an "Outstanding" from "Very Good" state.

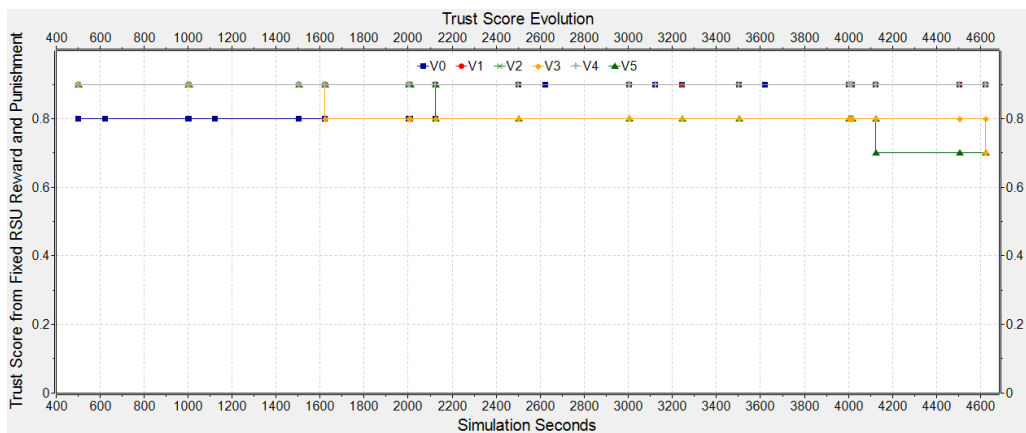

**Figure 18:** Trust score changes of sender and reporter drivers

Reporter drivers are untrustworthy throughout the experiment. This is happening as it is demonstrating a fact that trustworthy vehicles from higher-trust states can report fraudulently. As V0's announcement was always true, all clarifiers support this fact to the resolver RSU. Figure 18 shows this scenario, where simulation seconds are shown on the x-axis and trust changes are shown on the y-axis.

*With a Trust Value of 0.7 for all Drivers*

In this experiment, all vehicles start their journey with 0.7 from a "Fairly Good" trust state. The first announcement is reported by V2, and V2 receives RSU punishment at 620s and then enters into "Above Normal" state. V0 receives the RSU reward at 1100s and enters into "Good" trust state. Later, at 2000s the announcement is reported by V5 for which V0 receives the RSU reward and enters into the "Very Good" state at 2120s. V5 receives RSU punishment at 3100s and enters into "Above Normal" state. V0 enters into "Outstanding" trust state. Then the announcement at 3500s is reported by V4 dishonestly and receives RSU punishment at 3620s enters into "Very Good" state. V0's trust is not changed as it reaches the maximum trust of the framework. V0's next announcement is reported by V5 at 4500s, and the reporting from V5 is proven false to an RSU. Hence, it receives RSU punishment at 4600s and enters into "Normal" trust state. V0's trust is not updated with this reward, as V0 already hits the peak trust. Figure 19 shows this scenario, where simulation seconds are shown on the x-axis and trust changes are shown on the y-axis.

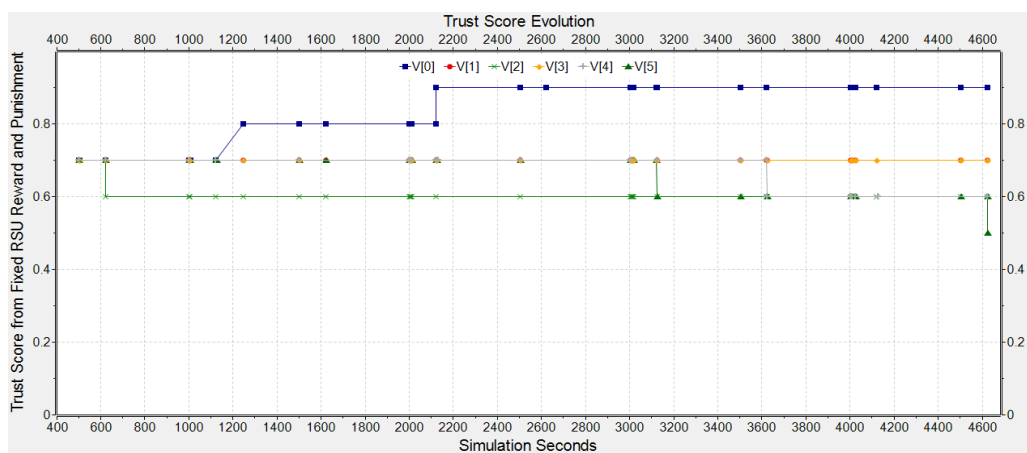

**Figure 19**: Trust score changes of sender and reporter drivers

*With a Trust Value of 0.7 for the Sender Driver and 0.8 for all Drivers*

Figure 20 shows similar characteristics to Figure 19. V0 shows trustworthy behaviour always. V0 starts from "Fairly Good", and reporters start from "Good" trust state. The announcements of V0 are reported at 1000s, 1500s, 2500s, 3000s, and at 4000s. In all cases, V0 win the disputes, but all reporters lose trust. Thus, V0 moves from "Fairly Good" to "Good" and then to "Outstanding". This chart shows that when V0 reaches peak trust, its curve becomes stable, but some reporters show trust decline in their curves as they generate false reports. For instance, V3 moves from "Good" to "Fairly Good" and then to "Above Normal". After that, V3 enters the "Normal" state. In this experiment, many trust state changes were seen in the sender and reporter drivers. Figure 20 shows this scenario, where simulation seconds are shown on the x-axis and trust changes are shown on the y-axis.

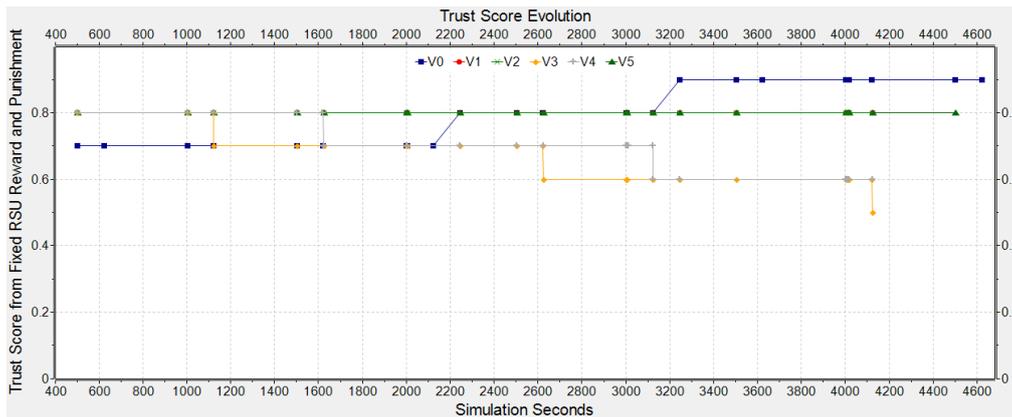

**Figure 20**: Trust score changes of sender and reporter drivers

*With a Trust Value of 0.6 for the Sender Driver and 0.7 for all Drivers*

Figure 21 shows the trust characteristics curve for the sender and reporter vehicles when the sender vehicle starts with 0.6 ("Above Normal" state) and all reporter vehicles start with 0.7 ("Good") trust scores. The first announcement is trustworthy, but the 2$^{nd}$ announcement from V0 is reported by V1 at 1000s, which is not a true report, so V1 receives RSU punishment. V1 moves from "Good" to "Above Normal". V0 enters into "Good" trust state. V3 receives punishment for the announcement at 1500s. V0 receives RSU reward at 2200s and 3300s for these two announcements from the resolver RSU. Thus, V0 enters into "Very Good" and then to the "Outstanding" state. Then the announcements at 2500s and 3000s are reported by V3, for which it receives RSU punishment, but V0 receives RSU reward for the first reporting by V3 at 3240s. V0's trust is not changed after it touches the peak trust, and it wins against the reporter vehicles later in all disputes. In this case, V3 enters the "Normal" state and then switches into the "Bad" state, from where it cannot report any announcement as the trust framework does not support the announcement of an accident message from this state. Figure 21 shows this scenario, where simulation seconds are shown on the x-axis and trust changes are shown on the y-axis.

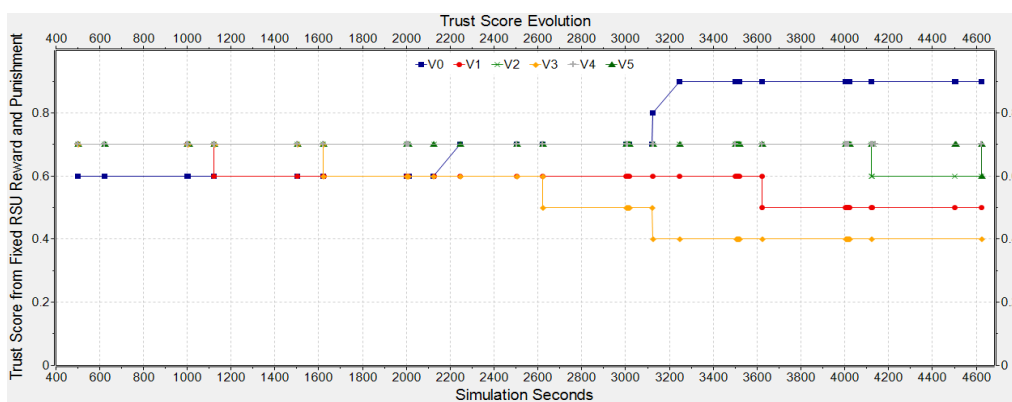

**Figure 21**: Trust score changes of sender and reporter drivers

*With a Trust Value of 0.6 for all Drivers*

This experiment illustrates that vehicles start their journey with a 0.6 trust score from "Above Normal" trust state. The first announcement from V0 is reported falsely by V1. V1 receives punishment at 620s, and V0 receives a reward from this reporting at 1100s. Thus,

V0 enters "Good" trust state, and V1 enters "Normal" trust state. V3 consecutively reports announcements from V0 at 1500s and 2500s, which are proven to be false to an RSU, so it receives RSU punishment at 1620s and 2620s. Thus, V3 moves into "Bad" trust state from "Above Normal" trust state. Alternatively, V0 improves trust from these reports and reaches "Outstanding" trust state by traversing "Very Good" trust state. The announcements of V0 at 3000s and 3500s are reported by V2 and V1 consecutively, and they are proven false to an RSU. So, their trust is reduced, but V0's trust remains the same, as it has already hit the peak. In this way, both drivers enter into lower trust states by issuing false reports to an RSU. Figure 22 shows this scenario, where simulation seconds are shown on the x-axis and trust changes are shown on the y-axis.

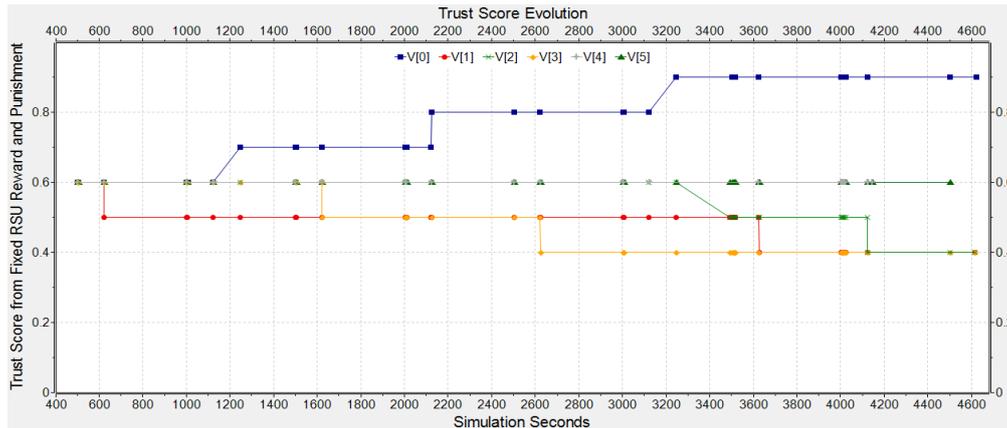

**Figure 22**: Trust score changes of sender and reporter drivers

*With a Trust Value of 0.5 for the Sender Driver and 0.6 for all Drivers*

This experiment shows the trust changes of sender and reporter drivers, where the sender driver starts with a 0.5 trust score from "Normal" trust state, and the reporter drivers start with 0.6 trust scores from "Above Normal" trust state. These drivers start with the same trust scores, but they finish with different trust scores, which depend on their trustworthiness throughout their journey. V1 reports an announcement of V0 at 500s for which it receives punishment and moves into "Normal" trust state from "Above Normal" trust state. V0 receives the RSU reward later at 1235s and enters into "Above Normal" state. Another announcement from the sender is reported by V2, which receives RSU punishment at 1120s, and it also moves into "Normal" state, but V0 receives RSU reward at 2230s. In the simulation, reward and punishment messages are issued at different times to avoid the collision of reward and punishment messages. V0 now enters into "Good" trust state. V1 again reports V0's announcement, which is also false, and receives RSU punishment at 1620s and enters into "Bad" trust state. V0 receives the RSU reward later at 3100s and moves to "Very Good" trust state. The announcement of V0 is reported by V3 at 2500s, receives RSU punishment at 2620s, and moves into "Normal" state. V0 receives the RSU reward at 3250s and enters into the "Outstanding" state. At this point, V0 reaches the peak trust, so most of its announcements are accurate. Later, the reporters V2 and V5 become fraudulent, and they receive RSU punishments. V0 receives RSU rewards from these disputes, and these are not added to V0's trust, as it already reaches the highest trust. V2 enters into "Below Normal" and then "Fairly Bad". This simulation shows both trust building and trust loss. It shows the transition of trust states from "Normal" to "Outstanding" and also the negative attitude of V2, which finishes with a "Fairly Bad" trust state, though it starts with a "Normal" trust state. This experiment catches the advantages of using additional trust states in the driver behaviour model and shows both positive and negative attitudes of drivers. Figure 23 shows this scenario, where simulation seconds are shown on the x-axis and trust changes are shown on the y-axis.

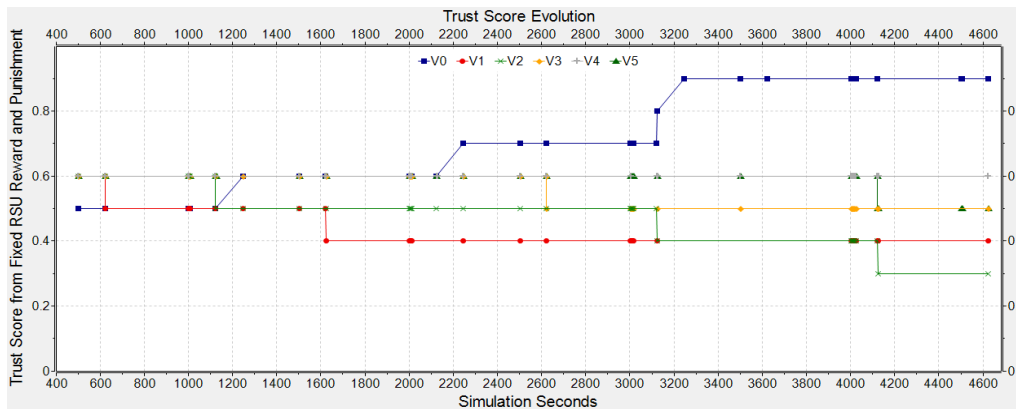

**Figure 23**: Trust score changes of sender and reporter drivers

## DISCUSSION ON RESULTS

In this set of experiments, a refined driver behaviour model incorporating eleven distinct trust states is implemented. Analysis of the experimental data reveals that this higher-resolution model allows for a more dynamic representation of trust, with drivers switching between states more frequently than in previous experiments using seven-state models. The utilization of eleven trust states provides significant advantages in terms of granularity, allowing for a precise, nuanced differentiation of driver trustworthiness. Specifically, within the higher range of trust scores, where a seven-state model might aggregate behaviours into broad categories (e.g., "Good," "Very Good," "Outstanding"), the eleven-state model enables a more detailed classification—such as distinguishing between "Above Normal", "Fairly Good", "Good", "Very Good", and "Outstanding". This increased granularity ensures that the model can capture subtle transitions in driver behaviour that would otherwise be lost. By providing this higher level of detail, I can classify, analyse, and predict driver trust with greater accuracy than previous models, enabling better calibration of automation to prevent over-trust or distrust.

## CONCLUSION

In this paper, driver announcement and reporting trustworthiness behaviour is analysed using three different Markov chain-based state transition diagrams. These three models have a different number of states: four, seven, and eleven. After examining these three models regarding driver announcement behaviour, the results suggest that the model with eleven states captures more dynamic behaviour for both announcement and reporting activities. Drivers switch more frequently among different trust states, both back and forth, according to the probability distribution.

# REFERENCES

1. Shahariar, R. and Phillips, C., 2024. The need for trustworthy announcements to achieve driving comfort, International Journal of Wireless & Mobile Networks (IJWMN), Vol.16, No.1/2, pp. 13-25

2. Shahariar, R. and Phillips, C., A Traffic INCIDENT MANAGEMENT FRAMEWORK FOR VEHICULAR AD HOC NETWORKS, International Journal of Wireless & Mobile Networks (IJWMN), Vol.17, No. 2, pp. 19-41

3. Shahariar, R. and Phillips, C. (2025) 'A Survey of Security Threats and Trust Management in Vehicular Ad Hoc Networks', *TECS*, 13, pp. 127-172.

4. Shahariar, R. and Phillips, C., 2024. A trust management framework for vehicular ad hoc networks, International Journal of Security, Privacy and Trust Management (IJSPTM) Vol 12, No 1, pp. 15-36

5. Shahariar, R. and Phillips, C., 2024. A fuzzy reward and punishment scheme for vehicular ad hoc networks, International Journal of Advanced Computer Science and Applications (IJACSA), Vol. 14, No. 6, pp. 1-17

6. Bhatt, D. and Gite, S., 2016, March. Novel driver behaviour model analysis using hidden Markov model to increase road safety in smart cities. In *Proceedings of the Second International Conference on Information and Communication Technology for Competitive Strategies* (pp. 1-6).

7. Zaky, A.B., Khamis, M.A. and Gomaa, W., 2022. Markov Switching Model for Driver Behavior Prediction: Use Cases on Smartphones. In *Recent Innovations in Artificial Intelligence and Smart Applications* (pp. 255-275). Cham: Springer International Publishing.

8. Cojocaru, I., Popescu, P.S. and Mihaescu, M.C., 2022, October. Driver Behaviour Analysis based on Deep Learning Algorithms. In *RoCHI* (pp. 108-114).

9. Peppes, N., Alexakis, T., Adamopoulou, E. and Demestichas, K., 2021. Driving behaviour analysis using machine and deep learning methods for continuous streams of vehicular data. *Sensors*, *21*(14), p.4704.

10. Negash, N.M. and Yang, J., 2023. Driver behavior modeling toward autonomous vehicles: Comprehensive review. *IEEE Access*, *11*, pp.22788-22821.

11. Tangade, S., Manvi, S.S. and Hassan, S., 2019, September. A deep learning based driver classification and trust computation in VANETs. In *2019 IEEE 90th Vehicular Technology Conference (VTC2019-Fall)* (pp. 1-6). IEEE.

12. Zhang, R., Tan, Z., Lin, Z., Zhang, R. and Liu, C., 2025. Exploring the trust and behavior of experienced advanced driver assistance system drivers: An on-road study. *Accident Analysis & Prevention*, *217*, p.108071.

13. Malik, M., Nandal, R., Dalal, S. and Jalglan, V., 2022. Deriving Driver Behavioral Pattern Analysis and Performance Using Neural Network Approaches. *Intelligent Automation & Soft Computing*, *32*(1).

14. Sommer, C., German, R. and Dressler, F., 2010. Bidirectionally coupled network and road traffic simulation for improved IVC analysis. *IEEE Transactions on mobile computing*, Vol. 10, No. 1, pp.3-15.

**Authors' Profiles**

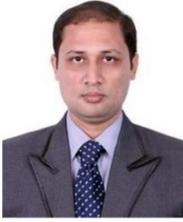

Rezvi Shahariar obtained his B.Sc. degree and an M.S. degree in Computer Science and Engineering from the University of Dhaka, Bangladesh in 2006. After some time working as a lecturer at the University of Asia Pacific, Dhaka, Bangladesh, he is now an Associate Professor at the Institute of Information Technology, University of Dhaka. Previously, he worked as an Assistant Professor and Lecturer at the same institute. He received a PhD on trust management for vehicular ad hoc networks at Queen Mary, University of London. His research focuses effective resource management in wireless network environments with an emphasis on trust, security in VANETs, fuzzy modelling, driver behaviour modelling, and the application of machine learning to security